\documentstyle[twoside,fleqn,psfig,espcrc2]{article}

\newcommand{\be}{\begin{equation}}
\newcommand{\ee}{\end{equation}}
\newcommand{\bea}{\begin{eqnarray}}
\newcommand{\eea}{\end{eqnarray}}

\newskip\humongous \humongous=0pt plus 1000pt minus 1000pt
\def\caja{\mathsurround=0pt}
\def\eqalign#1{\,\vcenter{\openup1\jot \caja
	\ialign{\strut \hfil$\displaystyle{##}$&$
	\displaystyle{{}##}$\hfil\crcr#1\crcr}}\,}
\newif\ifdtup

\relax

\newcommand{\AmS}{{\protect\the\textfont2
  A\kern-.1667em\lower.5ex\hbox{M}\kern-.125emS}}

\title{Heterotic versus Type I}

\author{C. Bachas\address{Centre de Physique Th\'eorique,
        Ecole Polytechnique, 
        91128 Palaiseau, France}%
        \thanks{Talk given at  STRINGS `97, Amsterdam,
June 16-21,  and at HEP-97, Jerusalem, August 19-26 }}

\begin{document}

\begin{abstract}
I compare the (prototype) calculations of special $F^4$ and $R^4$ 
 terms in the effective
actions of the $Spin(32)/Z_2$  heterotic and type I string
 theories, compactified on $S^1$ and ${\cal T}^2$.
 Besides checking duality, this elucidates  the quantitative
rules of  D-brane calculus.
 I explain in particular (a) why D-branes do
not run in loops, and (b) how their instanton contributions
arise from orbifold fixed points of   their  moduli space. 
\end{abstract}

\maketitle

\section{Introduction}

  The conjectured duality \cite{Witt,Chris,Dab,Dstr}
 between the type I and heterotic
$Spin(32)/Z_2$ string theories is particularly intriguing. 
The massless spectrum of both theories, in 10  space-time dimensions,
contains the (super)graviton and the  (super)Yang Mills multiplets.
Supersymmetry and anomaly cancellation 
fix  completely  the  low-energy Lagrangian, and more precisely
all two-derivative terms  and  the anomaly-cancelling,
four-derivative Green-Schwarz couplings
\cite{GSW,Ark}. One logical possibility, consistent with this
unique low-energy behaviour, could have been that the two theories are
{\it self-dual}  at strong coupling.  The conjecture that they
are instead dual to each other implies  that this unique infrared
physics  also has a unique consistent 
ultraviolet extrapolation.

    One of the early arguments in favour of this duality \cite{Chris,Dab}
was that the heterotic string appears as a singular solution of the
type I theory. Strictly-speaking this is not  an  independent
test of duality.  The presence of  a
 heterotic-string source will excite the
massless heterotic  backgrounds, and since the effective Lagrangian
is unique we should  not be surprised to find  the same solution 
on the type-I side. The real issue is whether consistency of the
theory {\it forces} us to include such excitations in the spectrum.
This can for  instance be argued in the case of  type
 II string theory near a
conifold singularity of the Calabi-Yau moduli space \cite{con}. 
Interestingly enough it can also be argued  for  the
simplest field-theoretic analog,  the  Dirac monopole 
of an abelian  theory in four dimensions. To render the 
ultraviolet theory  consistent, one must embedd the photon field
into  a spontaneously-broken,  asymptotically-free gauge model.
The point-like monopoles are then promoted to smooth solitons, which can
be pair-created and must thus  be included.

I am not aware of such a direct argument in the case of the heterotic
string solution. What is, however, known is that it
has  an exact conformal description
as a D(irichlet) string of  type I theory \cite{Dstr}.
In certain  ways
D-branes lie between fundamental quanta and smooth solitons
\cite{Joe,D} so, even if we admit  that they  are
intrinsic,  we must still decide on the rules for
including  them  in a semiclassical  calculation.
Do  they contribute, for instance, to loops like
fundamental quanta?  And with what measure and degeneracy should we
weight  their Euclidean trajectories?
In this talk I will review a prototype calculation \cite{BaKi,BFKOV,P}
in which these questions can be  answered.
Some  related calculations
in open string theory  can be found in refs. \cite{APT,SM,KO}.
The rules consistent with  duality turn out to be  natural and
simple.   D-strings, like smooth  solitons, 
 do not enter {\it explicitly} in loops
\footnote{A (light) soliton loop can of course
 be a useful approximation
to the exact instanton sum, as is the case  near the strong-coupling
singularities of the Seiberg-Witten solution \cite{SW}.},
while their  (wrapped)  Euclidean trajectories contribute to
the saddle-point sum, without topological degeneracy if one takes into
account correctly the non-abelian structure of D-branes. 
The way in which
 the various pieces of the  calculation fall
in place is, I believe,  further evidence
for  an underlying unique and elegant structure.


\section{BPS states and the unfolding trick}

   The prototype calculation is that of  $F^4$ and $R^4$  terms
in the effective quantum action, 
after torroidal compactification to $d>4$ dimensions.
Except  for  a particular CP-even combination \cite{Ark},
 all  these terms 
are special for the following reasons: (a) they are non-trivial,
since supersymmetry and anomaly cancellation do not fix them  completely
below 10  dimensions, and  (b) they 
are believed to be sensitive only to the BPS sector of the theory,
meaning that they are only corrected by  short multiplets at one loop,
 and by maximally-supersymmetric saddle points.
In these respects they are analogous to
 $F^2$ and $R^2$ terms in  vacua  with 8
unbroken supercharges \footnote{The non-renormalization statements
stand however  on less firm a ground, see for instance \cite{DS}.}.

The only (known)  supersymmetric saddle points on the heterotic side
are Euclidean trajectories of solitonic  five-branes. 
Since for  $d>4$ non-compact dimensions  these have no
finite 6-cycle to wrap around,  we expect the heterotic
one-loop result to be exact. 
For  zero Wilson lines, this one-loop amplitude  reads
\cite{Schellekens,Lerche}
\begin{equation}
\eqalign{
{\cal I}^{\rm heter}_{\rm 1-loop} &  =
-{V^{(d)}\over 2^{10} \pi^6}\
  \int_{\rm { Fund \atop dom}}{d^2\tau \over
\tau_2^2}\; (2\pi^2 \tau_2)^{5-d/2} \times  \cr &\times 
  \Gamma_{10-d,10-d}\
{\hat {\cal A}}({ F,R}, \tau) \cr}
\label{5}
\end{equation}
where  $V^{(d)}$ is the volume of the non-compact space-time,
$\Gamma_{10-d,10-d}$ is the usual  sum over momenta
 and windings on the
compactification torus \cite{GSW}, and
${\hat {\cal A}}$ is an  (almost) holomorphic modular
 form of weight zero,
closely related to the elliptic genus \cite{Paul}. More precisely,
 the Lagrangian form
 of the lattice sum is
\begin{equation}
\eqalign{ \ \ \ \ \   \Gamma_{10-d,10-d}
 &= \Bigl({2\over \tau_2}\Bigr)^{5-d/2}
\sqrt{\det G}\  \times  \cr   \times  \sum_{n^i,m^i}
&e^{-{2\pi\over\tau_2}
 (G+B)_{ij} ( m^i \tau-n^i) ( m^j \bar\tau-n^j)} \cr}
\end{equation}
with $G_{ij}$ the metric and $B_{ij}$ the (constant)
antisymmetric-tensor background on the torus.
I  use the  conventions  $\alpha' = \frac{1}{2}$,  and 
$G= L^2$ for a circle with radius $L$. Finally,
the elliptic genus ${\cal A}$  is
a chiral partition function, 
with extra weights for gauge-charge,  R-charge and spin
operators,  within the  Cartan subalgebra of
 SO(32)$\times$SO(10-d)$\times$SO(d-2). 
Expanding it out to fourth order in the charges and/or spins, and
regularizing in a modular-invariant way, yields ${\hat{\cal A}}$
\cite{Schellekens,Lerche}.

 This  heterotic amplitude can be most easily derived in the 
(light-cone) Green-Schwarz formulation \cite{Lerche}.
Let me concentrate for definiteness  on the gauge part of the effective
action.  The coupling of the heterotic string
to a constant (transverse) field-strength background reads
$$
\delta I_{\sigma} \  \propto\  \int \;   F_{ij}^\alpha  J_{\alpha}\ 
( X^i {\bar \partial} X^j - {1\over 8}  S^\dagger  \gamma^{ij} S )
$$
with $S$ the  Green-Schwarz fermions,  and  $J_\alpha$ the SO(32) 
world-sheet currents that can be represented as fermion bilinears,
$J_\alpha = T_\alpha^{rs} \lambda_r\lambda_s$.
  Consider now the $\sigma$-model functional
integral on the torus. To absorb the eight fermionic zero modes  
we must bring down at least  four powers of the fermionic piece of
$\delta I_{\sigma}$. The result is  proportional to the (covariantized)
eight-index tensor
$$
t_{(8)}^{i_1j_1 ... i_4j_4} =  \int [{\cal D} S_0] \; 
   (S_0^\dagger  \gamma^{i_1j_1} S_0)\;  ...
\;  (S_0^\dagger  \gamma^{i_4 j_4} S_0) \ ,
$$
times the momentum and winding  sum, times  the partition function of
left-moving (holomorphic)  states with four insertions of the SO(32)
generators. Notice that the bosonic and fermionic
 determinants cancel out on
the right-moving (supersymmetric) side.
Putting all this together  leads to  expression (\ref{5}) with
\begin{equation}
\eqalign{ 
{\hat {\cal A}}(F,\tau)& =  t_{(8)}
 {\rm tr}{ F}^4 
+{1\over 2^9\cdot 3^2}\; \Bigl[  {E_4^3\over \eta^{24}} +
 {\hat E^2_2 E_4^2\over \eta^{24}} \cr &
-2 {\hat E_2E_4E_6\over \eta^{24}} -2^7\cdot
3^2\Bigr]\   t_{(8)}\; ( {\rm tr} { F}^2)^2 \ ,  \cr}
\label{genus}
\end{equation}
where the  traces are  in the fundamental representation of SO(32)
and I have suppressed Lorentz indices. 
The Eisenstein series $E_{2k}(q=e^{2\pi i\tau})$ are 
modular forms of weight $2k$. They are holomorphic with
the exception of $\hat E_2$,
which  requires  a non-holomorphic regularization.

   In its  form (\ref{5})  the amplitude is manifestly sensitive
only to  heterotic BPS states. Such  states  are characterized by
the fact that there are no oscillator excitations in the right-moving
sector. The same type of argument can be applied to the open superstring
\footnote{BPS saturation of the $F^4$ terms was
first noticed in \cite{DKPS},  and established through helicity
supertrace formulae in \cite{BaKi}.} .
Since the left- and right-moving sectors are
 not independent in this case,
the only BPS states are the
(Kaluza-Klein descendants) of the  SO(32) (super)gauge bosons. 
After Poisson resummation of the Kaluza-Klein momenta we get
\begin{equation}
\eqalign{ 
{\cal I}^{\rm open}_{\rm 1-loop}  = 
 -{V^{(10)}\over
  2^{10}\pi^6}& \ \  t_{(8)} {\rm tr}_{\rm adj} { F}^4 \times \cr  
 \times  \int_0^\infty & {dt\over t^2} 
\sum_{n^i\not= \{0\}} 
 e^{- 2\pi G_{ij} n^i n^j/ t} \cr}
\label{I}   
\end{equation}
with the trace in the adjoint representation of SO(32). 
The quadratically-divergent $n^i = \{0\}$ term has been subtracted
explicitly from the sum. A carefull calculation \cite{BF}
shows that it indeed corresponds to a (one-particle-reducible)
diagram, with a massless graviton exchanged in the transverse
channel. This is the only way in which (\ref{I}) differs from
the naive (super)Yang Mills expression.

Consider now a circle compactification to 9 dimensions.
Since there are no stable 2-cycles around which Euclidean D-string
trajectories may  wrap, we  expect no instanton corrections
on the type I side.  Hence  the action (\ref{I}) should be (almost)
exact. But this
 raises an obvious puzzle: 
 The heterotic spectrum contains
an infinite tower of BPS states  in arbitrary representations of
$Spin(32)/Z_2$, all of which contribute to the one-loop integrand
in (\ref{5}). On the type I side these   correspond to  states of a 
 D-string,   winding
around the compactification  circle.
If treated as  regular solitons,
D-strings should not enter  explicitly in loops.
With so  many more
states ``running'' in the heterotic as compared to  the type-I loop, 
how can the two expressions  possibly match?

The answer to this puzzle makes use of an old trick
familiar from the study of free-string thermodynamics \cite{Mc}.
The idea is to trade part of the lattice sum, so as to unfold the
integration region from a fundamental domain into the strip,
$$
(2\pi^2\tau_2)^{1\over 2}  \Gamma_{1,1}
 = 2\pi L  \  \Bigl[\  1+ 
 \sum_{n\not= 0 \atop   {\cal S}\in\; {\rm strip} }
 e^{- 2\pi   L^2 n^2 /{\rm Im} {\cal S}(\tau)} \Bigr] , 
$$
where ${\cal S}$ labels  all modular transformations that leave
$\tau$ inside the  strip $-{1\over 2} \leq \tau_2 < {1\over 2}$. 
Using the modular invariance of 
 ${\hat {\cal A}}$, we find
\begin{equation}
\eqalign{
{\cal I}^{\rm heter}_{\rm 1-loop}& = 
 -{V^{(10)} \over 2^{10} \pi^6}\; \; 
\Bigl[  \int_{{\rm Fund\atop dom}}{d^2\tau \over
\tau_2^2}\; + 
 \cr 
+& \int_{\rm strip} {d^2\tau\over\tau_2^{\ 2}}\;
 \sum_{n \neq 0}
 e^{- 2\pi L^2
n^2/\tau_2 }\; \Bigr]\;{\hat {\cal A}}(F,\tau) \ .
\cr}
\label{trick}
\end{equation} 
The  $\tau_1$ integration in the strip  kills all but the $q^0$
piece of the
 elliptic genus,
\begin{equation}
\eqalign{ 
\ \ \ {\hat {\cal A}} \bigg|_{q^0} \  =\
&  t_{(8)}\; {\rm tr}_{\rm adjoint} F^4 - \cr
- & \Bigl[{15\over 16\pi\tau_2}  
-{63\over 64\pi^2\tau_2^2} \Bigr]\  t_{(8)}\;  ({\rm tr}F^2)^2 \ .\cr} 
\end{equation}
Plugging  the first of these terms inside (\ref{trick}) gives
precisely  the one-loop type I expression.
{\it The massive BPS states conspired with the stringy cutoff
 on the heterotic side, to reproduce the simple  loop of  SO(32)
gauge bosons}.

  The heterotic one-loop  contains in fact extra terms, besides the
expression (\ref{I}). They  can be organized as
 an expansion  in inverse powers
of the radius. Since the latter  gets rescaled
  by duality,  $L^2 \rightarrow 
L^2/g_s$  with $g_s$ the string coupling constant, these  extra
contributions  must   come from diagrams of genus $\not= 1$ on the type I
side. The leading term in the decompactification limit is given by the
integral of ${\hat {\cal A}}$  over a   fundamental domain. It
is equal to   the quartic piece  of the Born-Infeld action, which 
arises from  the type I disk diagram \cite{Ark}. The 
subleading term is the one-loop contribution. 
The two remaining  terms come from  the non-holomorphic regularization
of ${\hat {\cal A}}$. They correspond to  contact
 contributions  in two- and
three-loop open string diagrams. It is conceivable that
matching all lower-dimensional
operators in the effective heterotic and type I actions requires
 field redefinitions which absorb these terms \cite{APT}.


\section{ D-brane instantons}

   Let us now move one step further down  and  consider a
${\cal T}^2$  compactification  of the  eighth and ninth spatial
dimensions. The lattice
 sum on the heterotic side
takes  the form
\begin{equation}
  {\Gamma_{2,2}}  =
{ T_2\over \tau_2 }
 \sum_{ M }
  e^{ 2\pi i  T\  {\rm det}M 
- \frac{\pi T_2 }{ \tau_2 U_2 }
\big| (1\; U)M  \big( {\tau \atop -1} \big) \big| ^2 } 
\label{DKL}
\end{equation}
where $M$ runs over all  $2\times 2$ matrices  with integer entries,
and  
$$
 U = ( G_{89} + i\sqrt{G} )/ G_{88} \ \ {\rm and}\ \ 
 T= \frac{1}{\alpha' } (B_{89} + i\sqrt{G})\ 
$$
are the complex structure and K\"ahler moduli of ${\cal T}^2$.
The matrix $M$ describes the wrapping of the  heterotic
world sheet 
around the target-space  torus. The two
generators of the world-sheet torus are given, as  vectors 
in the compactification lattice, 
 by the columns of the matrix $M$.
The exponent in equation (\ref{DKL})
 is the minimum Polyakov action for a given wrapping.

The contributions of degenerate matrices ($det M = 0$)  
sum up to  the perturbative type I result. This  follows
 from an argument
identical to the one used in  the previous section.
 The novel feature are
non-degenerate matrices,  which  correspond to
 heterotic world-sheet instanton
corrections. Using a global  world-sheet reparametrization,
 one can bring
such a  matrix $M$ to  the canonical form
$$
M =  \pm \left(\matrix{ k& j\cr 0&p\cr}\right) \ \ 
{\rm with}\ \   0\leq  j <k \  ,   \ \  p\not= 0\ .
$$
The sum over the PSL(2,Z) orbits of these matrices can furthermore
be traded against unfolding the fundamental domain integral  onto
(twice)  the  upper complex plane \cite{DKL}. Performing  explicitly
the integral leads to the following expression for the one-loop
heterotic action \cite{BFKOV}
$$
 {\cal I}^{\rm heter}_{\rm 1-loop} = {\cal I}_{\rm degen}  + 
 {\cal I}_{\rm inst}
$$
where 
\begin{equation}
\eqalign{
{\cal I}_{\rm inst}= 
 -{ V^{(10)}\over 2^{9}\pi^6} &
  \sum_{{0 \leq j<k} \atop { p > 0}}
      { e^{2\pi i pk T} \over k p T_2}\; \times \cr
  & \times  {\cal O}  \;  {\hat {\cal A}}\left({j+p U\over k}\right)
 +\  {\rm c.c.}
\cr} 
\label{inst}
\end{equation}
Here  ${\cal O} = 1 + ...$ is a differential
 operator,  whose action is non-trivial
only because of the non-holomorphicities of the elliptic genus. 
I will ignore this complication  by assuming for instance
that ${\rm tr}F^2 = {\rm tr}R^2$,  in which case ${\hat {\cal A}}$
is completely holomorphic.

\begin{figure}
%
\centerline{\psfig{figure=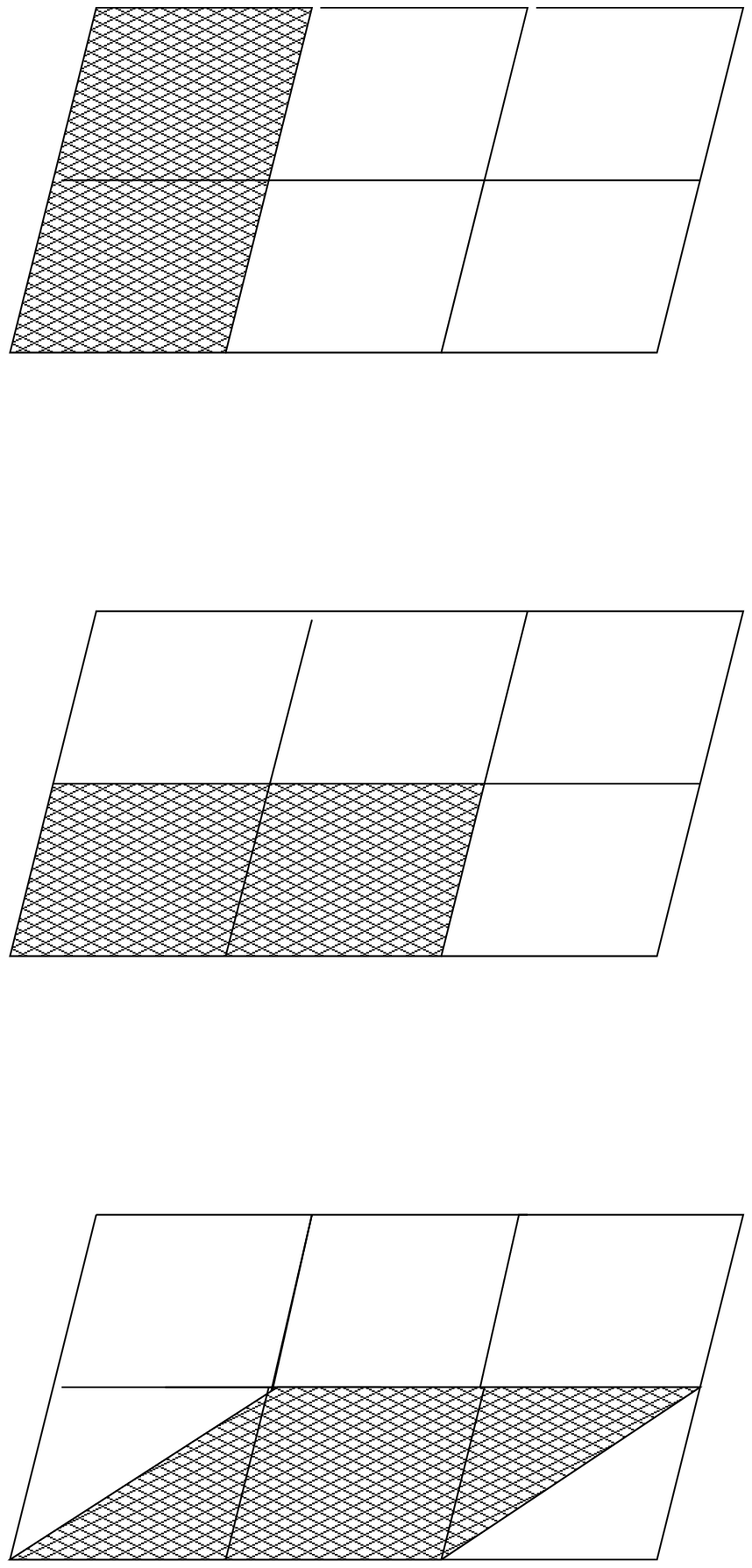,height=9cm}}
%
%
\caption{The three distinct ways in which the
(shaded)  heterotic world sheet
can wrap twice around the target-space torus.
\label{fig:radish}}
\end{figure}

 Expression (\ref{inst}) is a sum over all distinct ways to wrap
the target-space torus with the heterotic,  or D-string (Euclidean)
world sheet. The induced K\"ahler and complex moduli  on this
latter, for positive $p$ \footnote{Negative $p$'s 
correspond to antiinstantons, and give  complex-conjugated
expressions.},  read
$$
{\tilde T}  =  kp T \ \ \ {\rm and }\ \ \ 
{\tilde U}= {j+pU\over k}\  
 .
$$
The  three distinct ways to wrap the torus twice are, for instance,
drawn  in figure 1. They correspond   (from top
to bottom) to 
$$
M = \left( \matrix{1 & 0 \cr 0 & 2 \cr}\right) \ , \
 \left(\matrix{2 & 0 \cr 0 & 1 \cr}\right)\ \ {\rm and}  \
 \left(\matrix{2 & 1 \cr 0 & 1 \cr}\right)\ ,
$$
have  complex structures 
$${\tilde U}\  =\  2U\ ,\ \ 
 {U\over 2}\ \ {\rm  and} \ \  {1+ U\over 2}\ ,
$$
and (common) K\"ahler modulus  ${\tilde T} = 2T$.  
The   weight of    each surface in expression 
(\ref{inst}) can  be recognized easily  as the product of
 (a) the exponential of minus the  Nambu-Goto
action,   $I_{NG} = - 2\pi i {\tilde T} $, 
 (b)  the elliptic genus at  the induced
complex structure ${\tilde U}$, and (c) a factor of the inverse 
 area. All this is in accordance with the naive expectation that
 the auxiliary world-sheet metric may be eliminated   in favour of
a heterotic Nambu-Goto action.

  From the type-I point of view expression (\ref{inst}) is,
however, still somewhat unnatural.
 The three configurations of figure 1 correspond
to the same (singular) effective-field-theory solution, characterized
by two units of the appropriate Ramond-Ramond charge. Why then
should we count them as distinct saddle points?
 Furthermore, the fluctuations
of the double D-brane  are not described by the usual heterotic
$\sigma$-model, but by its (non-abelian)  $2\times 2$-matrix
generalization \cite{matrix},
which is the low-energy limit of an open string theory.
Why should then the result be  proportional to the conventional
elliptic genus?

In order to answer these questions \footnote{This argument has been
developped in collaboration with Pierre Vanhove \cite{P}.}
 it is convenient to put the
effective action (\ref{inst}) in the more elegant form
\begin{equation}
 {\cal I}^{\rm inst} =
- { V^{(8)}\over 2^{8}\pi^4}
\sum_{N=1}^\infty   \;  e^{2\pi i N T}
 {\cal H}_N  {\hat {\cal A}}(U)\ + {\rm c.c.},
\label{Hecke}
\end{equation}
with
\begin{equation}
{\cal H}_N {\hat {\cal A}}(U) \equiv
{1\over N} \sum_{kp = N \atop 0\le j <k}
 {\hat {\cal A}}\left({j+p U\over k}\right) \ .
\end{equation}
In the mathematics literature ${\cal H}_N$ is known as a Hecke
operator \cite{math}. We have just seen its geometric interpretation
in terms of inequivalent N-fold wrappings of the torus by the
(heterotic) world sheet. I will now describe an
 alternative interpretation,
more appropriate on the type I side, in terms of the moduli space of
instantons. The key 
 will be  to treat this moduli space as a
symmetric orbifold \cite{c,Cum,Vs}, an idea that is more familiar
in the context  of black-hole state counting \cite{BH}.

The low-energy
 fluctuations around  a configuration of
$N$ instantonic D-branes  are  described by a 
heterotic matrix $\sigma$-model, with local
SO(N) symmetry on the world sheet \cite{matrix}.
The
 coupling of  a
constant target-space  background field reads
$$
\delta I_{\sigma}  \  \propto\ 
 \int \;   F_{ij}^\alpha T_\alpha^{rs}\ 
\lambda_r^{\; T} \ 
\Bigl[  X^i {\bar D}  X^j - {1\over 8}  S^\dagger 
 \gamma^{ij} S \Bigr] \lambda_s \ .
$$
Under the SO(N) gauge symmetry the
supercoordinates $X^i$ and $S^{\dot a}$ are symmetric
matrices,  the  current-algebra fermions
$\lambda_r$  are vectors, and ${\bar D}$ is the antiholomorphic
covariant derivative.
We are interested in the functional integral of
 this $\sigma$-model,
with four insertions of $\delta I_{\sigma}$. Notice that 
 contributions of massive string modes are expected to cancel out
for this special amplitude, justifying  the reduction of the
calculation to the matrix model.

The moduli space of this multiinstanton has  a Higgs branch 
along which the $X^i$ have diagonal
expectation values. In the type $I^\prime$ language these label
the  positions of  N D0-particles on  the orientifold plane
\footnote{There is also a Coulomb branch, corresponding to the motion
of mirror pairs of D0-particles off the orientifold plane.
Because of the SO(N)-gaugino zero modes, this part of the moduli
space does  not contribute.}.
 At a generic point in this moduli space there are 8N fermionic
zero modes, corresponding to the diagonal components of the
matrices $S^{\dot a}$. Since only eight of them
can  be absorbed by the  insertions of the vertex $\delta I_{\sigma}$,
 one would naively conclude that the 
sectors $N>1$ do not contribute.  This is wrong because 
of the residual gauge symmetry that permutes the positions of the
D-branes. {\it The 
moduli space is thus a symmetric  orbifold  and there
are potential  contributions from its fixed points}.

  Let me illustrate how this works in  the case
 of two  instantons. The massless
fluctuations of the double D-brane are described by a conformal
field theory with target space ${\cal M}\times {\cal M}/Z_2$,
where ${\cal M}$ is the (transverse)
 target space of the heterotic string,
and $Z_2$ is the exchange symmetry. There are four contributions to the
amplitude, corresponding to the four 
boundary conditions on the torus. The untwisted sector has $2\times 8$
fermionic zero modes and does not contribute.
The contribution of
the remaining three sectors is proportional to
$$
{\hat {\cal A}}(2U) + {\hat {\cal A}}\Bigl({U\over 2}\Bigr)+
{\hat {\cal A}}\Bigl({U+1\over 2}\Bigr)\ ,
$$
as can be shown using  standard $Z_2$-orbifold techniques \cite{P}.
This is precisely the action of the Hecke operator ${\cal H}_2$,
corresponding to the sum over the three surfaces of figure 1. 
  The overall coefficient also
checks, including the orbifold normalization of ${1\over 2}$,  and
 the simple factor of the transverse volume
 characteristic of
the twisted-sector contributions.

  The generalization to any $N$ is straightforward. The target
space is now the symmetric orbifold 
$$
\underbrace{ {\cal M} \times ... \times {\cal M}}_{\rm N\  times}
\bigg/S_N 
$$
The non-vanishing contributions
to the amplitude come from those boundary conditions for which 
only the trace part of $S^{\dot a}$ is (doubly) periodic on the torus.
Up to a common overall normalization, the
 result is given by  ${\cal H}_N {\hat{\cal A}}(U)$,
which is
 the matrix-model generalization of  ${\hat{\cal A}}$.
The non-perturbative type I effective action is obtained by 
 summing  over all N,  as in  expression 
(\ref{Hecke}).

\section{Outlook}

 It would be very interesting to extend these considerations to
more involved settings, including in particular the type I D5 brane.
This should  lead to a simple combinatorial understanding, through
instanton calculus,  of the  Seiberg-Witten
solution.
\vskip 0.5cm

 I thank my collaborators C. Fabre, E. Kiritsis,  N. Obers and
especially P. Vanhove for many discussions. This research was 
partially supported by EEC grant  TMR-ERBFMRXCT96-0090.

\vskip 0.8cm 

\newcommand{\Journal}[4]{{#1} {\bf #2}, #3 (#4)}
\newcommand{\NCA}{\em Nuovo Cimento}
\newcommand{\NIM}{\em Nucl. Instrum. Methods}
\newcommand{\NIMA}{{\em Nucl. Instrum. Methods} A}
\newcommand{\NPB}{{\em Nucl. Phys.} B}
\newcommand{\PLB}{{\em Phys. Lett.}  B}
\newcommand{\PRL}{\em Phys. Rev. Lett.}
\newcommand{\PRD}{{\em Phys. Rev.} D}
\newcommand{\ZPC}{{\em Z. Phys.} C}

\end{document}